# Can a GPT4-Powered AI Agent Be a Good Enough Performance Attribution Analyst?

Bruno de Melo[1], Jamiel Sheikh

*CUNY – School of Professional Studies*
New York, NY, United States

**ABSTRACT**

Performance attribution analysis, defined as the process of explaining the drivers of the excess performance of an investment portfolio against a benchmark, stands as a significant feature of portfolio management and plays a crucial role in the investment decision-making process, particularly within the fund management industry. Rooted in a solid financial and mathematical framework, the importance and methodologies of this analytical technique are extensively documented across numerous academic research papers and books. The integration of large language models (LLMs) and AI agents marks a groundbreaking development in this field. These agents are designed to automate and enhance the performance attribution analysis by accurately calculating and analyzing portfolio performances against benchmarks. In this study, we introduce the application of an AI Agent for a variety of essential performance attribution tasks, including the analysis of performance drivers and utilizing LLMs as calculation engine for multi-level attribution analysis and question-answering (QA) tasks. Leveraging advanced prompt engineering techniques such as Chain-of-Thought (CoT) and Plan and Solve (PS), and employing a standard agent framework from LangChain, the research achieves promising results: it achieves accuracy rates exceeding 93% in analyzing performance drivers, attains 100% in multi-level attribution calculations, and surpasses 84% accuracy in QA exercises that simulate official examination standards. These findings affirm the impactful role of AI agents, prompt engineering and evaluation in advancing portfolio management processes, highlighting a significant development in the practical application and evaluation of Generative AI technologies within the domain.

## 1. Introduction

With the successful release of chatGPT in November 2022 and subsequent releases of similar open-source models, there has been increasing interest in applying large language models (LLM) to create AI systems or applications that are governed by AI agents. Theoretically, an AI agent can act, observe, generate tasks, and make specific decisions within a controlled environment in order to reach predetermined objectives or goals. Within the realm of LLMs, these agents can autonomously perform automation tasks such as code generation, personal assistance, information retrieval and summarization, etc. One of the main advantages of AI agents is the perceived increase in human productivity in that most of the mundane tasks could be automated, thus leaving us to think and do more creative tasks and jobs.

[1] Corresponding author. Email: bruno.verasdemelo27@spsmail.cuny.edu

Open-source LLM application wrappers like LangChain[2] and LlamaIndex[3] offer quick and easy-to-use pre-canned AI Agents that can generate meaningful results with just a few lines of codes. These functions can be customized and might serve as a jump start for more ready-to-production applications. There are other open-source frameworks that allows for customized agent creation like Ai Agent[4], AgentGPT[5], and Superagent[6], to name a few. A recent commercial application is Microsoft's 365 Copilot product, which integrates AI into its Office suite of products like Work, Excel, Teams. In Word, Copilot allows for draft creation, addition of new content to an existing document, rewriting of text, text summarization, etc.

There are various business processes that can benefit from the use of AI Agent, especially those processes that consume or produce structured data - which is commonly arranged in a way that is highly organized, easily manipulated, and understandable by machines, like tabular data. As opposed to unstructured data, typically found in books and research papers, structured data is typically found in relational databases and knowledge graph networks. Processes utilizing semi-structured data which serve as a bridge between structured and unstructured data, such as content produced in JSON or XML format, are also a clear beneficiary of the use of AI Agent.

Within the asset management industry, specifically in fund investing and portfolio management related activities, a business process that can be enhanced using AI agents relates to portfolio performance measurement and attribution. The measurement of portfolio performance versus a benchmark, or mostly commonly known as *performance attribution analysis*[7] is defined by Bacon (2008) as "a technique used to quantify the excess return of a portfolio against its benchmark into the active decisions of the investment decision process".

Performance attribution analysis is an integral and important part of the investment decision process, providing added value in that it allows for independent and emotion-free feedback into the investment process, and serving as a transparent way for senior management and clients to identify sources of portfolio returns and risks. Ideally, the performance attribution team should be part of a risk control function, never

---

[2] https://python.langchain.com/docs/modules/agents/
[3] https://llamahub.ai/
[4] https://aiagent.app/
[5] https://agentgpt.reworkd.ai/
[6] https://www.superagent.sh/
[7] Performance attribution analysis and attribution analysis will be used intertwined.

reporting to the front office or the marketing department, as recommended by Bacon (2008).

The performance attribution team employs specialized personnel with industry wide recognized certification programs such as the CIPM - Certificate in Investment Performance Measurement, managed by the CFA Institute[8]. Moreover, it is based on established financial and mathematical foundation with a plethora of published academic research papers and books. Many commercial software solutions widely used by the financial industry incorporate such techniques as part of their offering, such as Factset and MSCI's Barra.

Attribution analysis poses an interesting challenge to the use of an AI agent for several reasons, such as: i) it requires incorporating **domain specific knowledge** into LLMs, hence opening the possibility of **fine-tuning** and the use of **Retrieval Augmented Generation (RAG)** frameworks; ii) it involves reasoning on structured/tabular data which implies that LLMs have to be able to generate 100% accurate results, hence requiring the use of **hallucination controls** and **evaluation methods**; and iii) being a multi-dimensional, multi-temporal problem, it requires **LLMs to reason and to provide insights** using data from several columns before answering a query. We believe that an AI agent employed in this context could be a relatively cheap and complimentary tool in the toolbox of an experience performance attribution analyst in that it is a cost-effective way of reducing cognitive burden when analyzing multiple portfolios over multiple time scales, thus increasing their time to perform in-depth analysis, for example.

## 2.     Literature Review

This section introduces concepts and topics that are used throughout this research paper, such as performance attribution analysis, AI Agents, and evaluation methods.

Performance attribution – Defined as the measure of portfolio performance or excess return versus a benchmark "into the active decisions of the investment decision process", as per Bacon (2008). We can trace the origins of performance attribution analysis to a seminal study on performance measurement of pension funds in the UK from 1966, Dietz (1966). The main conclusions still resonate today:

   i. Funds should be classified according to their investment objectives.
   ii. Performance measurement returns should be based on market value, not cost.
   iii. Returns should be total returns and be time weighted.
   iv. Performance should include risk as well as return.

---

[8] https://www.cfainstitute.org/en/programs/cipm

The major development on attribution was done in the 1980s, especially on two papers: Brinson and Fachler (1985) and Brinson et al. (1986). These papers laid the foundation of what is commonly known today as the "Brinson model". The Brinson and Fachler (1985) paper is the first to document the formulas to calculate individual excess return attribution effects: Allocation, Selection, and Interaction. For an elucidative discussion on the history of the various formulas associated with the Brinson model, Bacon (2019) is a good reference.

From the various performance attribution calculation methods available, we use the Brinson-Fachler methodology in this research paper. Within this framework, considering a single period and single currency portfolio grouped by sectors, and assuming the portfolio follows a long-only strategy meaning it cannot sell short securities and it does not use leverage; implying that a portfolio sector weight cannot be negative, and the sum of weights is 100%.

Total Contribution or Total Effect for sector $i$ is defined as the sum of the allocation and selection effect for that sector, expressed as:

$$TC_i = A_i + S_i \qquad (2.1)$$

where: $A_i$ = allocation effect in the $i^{th}$ sector, and

$S_i$ = selection effect in the $i^{th}$ sector.

Let us also define the portfolio Total Contribution or Total Effect as,

$$TC = \text{Total Contribution} = \sum_{i=1}^{i=n} TC_i = \sum_{i=1}^{i=n}(A_i + S_i) = A + S \qquad (2.2)$$

where: $\sum_{i=1}^{i=n} A_i = A$ defines the portfolio total Allocation Effect and,

$\sum_{i=1}^{i=n} S_i = S$ defines the portfolio total Selection Effect.

From equation 2.2, the following possible cases are identified:

1) Total Contribution being positive:
    i. Allocation Effect and Selection Effect are both positive,
       OR.
    ii. Allocation Effect being positive and Selection Effect being negative, implies that Allocation is greater than the absolute value of Selection,
       OR
    iii. Allocation Effect being negative and Selection Effect being positive, implies that Selection is greater than the absolute value of Allocation.

2) Total Contribution being negative:
    iv. Allocation Effect and Selection Effect are both negative,

OR.

    v. Allocation Effect being positive and Selection Effect being negative, implies that Allocation is less than the absolute value of Selection,

OR

    vi. Allocation Effect being negative and Selection Effect being positive, implies that Selection is less than the absolute value of Allocation.

3) Total Contribution being zero:
    vii. Allocation Effect offset by the Selection effect.

Exhibit 1 summarizes the possible combinations.

|  | | Selection | | |
|---|---|---|---|---|
| | | $S_i > 0$ | $S_i < 0$ | $S_i = 0$ |
| Allocation | $A_i > 0$ | TC Positive | TC Positive or Negative | TC Positive |
| | $A_i < 0$ | TC Positive or Negative | TC Negative | TC Negative |
| | $A_i = 0$ | TC Positive | TC Negative | TC Zero |

Exhibit 1: Total Contribution Effect combination of factors

The Allocation effect for sector $i$ is expressed as:

$$A_i = (w_i - W_i) \times (b_i - b) \quad (2.3)$$

where: $w_i \geq 0$ = weight of the portfolio in the $i^{th}$ sector (note $\sum_{i=1}^{i=n} w_i = 1$),

$W_i \geq 0$ = weight of the benchmark in the $i^{th}$ sector, (note also $\sum_{i=1}^{i=n} W_i = 1$),

$b_i$ = return of the benchmark in the $i^{th}$ sector,

$$b = \text{Benchmark return} = \sum_{i=1}^{i=n} W_i \times b_i \quad (2.4)$$

Let us also define,

$$r = \text{Portfolio return} = \sum_{i=1}^{i=n} w_i \times r_i \quad (2.5)$$

where: $r_i$ = return of the portfolio of assets in the $i^{th}$ sector.

From equation 2.3, the following possible cases are identified:

1) Attribution effect being positive, implies that the first term and the second term must have the **same sign**:
    i. First term is positive: Portfolio sector weight is greater than Benchmark sector weight, or portfolio is overweight in that particular sector, AND
    ii. Second term is positive: Benchmark sector return is greater than the overall Benchmark return, or benchmark sector outperformed the overall market;
    OR
    iii. First term is negative: Portfolio sector weight is less than Benchmark sector weight, or portfolio is underweighting that particular sector, AND
    iv. Second term is negative: Benchmark sector return is less than the overall Benchmark return, or benchmark sector underperformed the overall market.

2) Attribution effect being negative, implies that first term and the second term must have **opposite signs**:
    v. First term is positive: Portfolio sector weight is greater than Benchmark sector weight, or portfolio is overweight in that particular sector, AND
    vi. Second term is negative: Benchmark sector return is less than the overall Benchmark return, or benchmark sector underperformed the overall market.
    OR
    vii. First term is negative: Portfolio sector weight is less than Benchmark sector weight, or portfolio is underweighted in that particular sector, AND
    viii. Second term is positive: Benchmark sector return is greater than the overall Benchmark return, or benchmark sector outperformed the overall market.

3) Attribution effect being zero, implies that either the first term is zero or the second term is zero or that both are zero:
    ix. First term is zero: Portfolio sector weight is the same as the Benchmark sector weight, or the portfolio is neutral weight in that particular sector.
    x. Second term is zero: Benchmark sector return is the same as the overall Benchmark return.

Exhibit 2 summarizes the possible combinations.

|  | Returns R | | |
|---|---|---|---|
|  | $(b_i - b) > 0$ | $(b_i - b) < 0$ | $(b_i - b) = 0$ |
| $(w_i - W_i) > 0$ | Allocation Positive<br>W: Overweight<br>R: Outperformance | Allocation Negative<br>W: Overweight<br>R: Underperformance | Zero Allocation<br>W: Overweight<br>R: Neutral |

| | Allocation Negative | Allocation Positive | Zero Allocation |
|---|---|---|---|
| $(w_i - W_i) < 0$ | W: Underweight<br>R: Outperformance | W: Underweight<br>R: Underperformance | W: Underweight<br>R: Neutral |
| $(w_i - W_i) = 0$ | Zero Allocation<br>W: Neutral<br>R: Outperformance | Zero Allocation<br>W: Neutral<br>R: Underperformance | Zero Allocation<br>W: Neutral<br>R: Neutral |

Exhibit 2: Allocation Effect combination of factors

The selection effect for sector $i$ is expressed as:

$$S_i = w_i \times (r_i - b_i) \qquad (2.6)$$

where: $w_i \geq 0$ = weight of the portfolio in the $i$th sector (note $\sum_{i=1}^{i=n} w_i = 1$),

$r_i$ = return of the portfolio of assets in the $i^{th}$ sector, and

$b_i$ = return of the benchmark in the $i^{th}$ sector.

From equation 2.6, the following possible cases are identified:

1) Selection effect being positive, assuming that Portfolio Weight cannot be negative, implies that term in brackets must be positive:
   i. Portfolio sector return is **greater** than the Benchmark sector return, or portfolio sector **outperformed** the benchmark sector.

2) Selection effect being negative, assuming that Portfolio Weight cannot be negative, implies that term in brackets must be negative:
   ii. Portfolio sector return is less than the Benchmark sector return, or portfolio sector underperformed the benchmark sector.

3) Selection effect being zero, implies that either the first term is zero or the second term is zero:
   iii. First term is zero: Portfolio sector weight is zero, or portfolio has no exposure to that sector,
   OR
   iv. Second term is zero: Portfolio sector return is the same as the Benchmark sector return.

Exhibit 3 summarizes the possible combinations.

| | Returns R | | |
|---|---|---|---|
| | $(r_i - b_i) > 0$ | $(r_i - b_i) < 0$ | $(r_i - b_i) = 0$ |

|  | | Selection Positive Outperformance | Selection Negative Underperformance | Zero Selection Neutral |
|---|---|---|---|---|
| Weights W | $w_i > 0$ | Selection Positive Outperformance | Selection Negative Underperformance | Zero Selection Neutral |
| | $w_i < 0$ | Not applicable | Not applicable | Not applicable |
| | $w_i = 0$ | Zero Selection No exposure | Zero Selection No exposure | Zero Selection No exposure |

Exhibit 3: Allocation Effect combination of factors

In the Brinson-Fachler methodology the Interaction Effect is added to the Selection Effect. As explained by Bacon (2019), "…combining selection with interaction allows the analyst to appropriately measure the impact of both the asset allocation and security selection decisions in the investment process."

The Brinson model with the Interaction effect added to the Selection effect is illustrated graphically in Exhibit 4 - extracted from Bacon and Wright (2019).

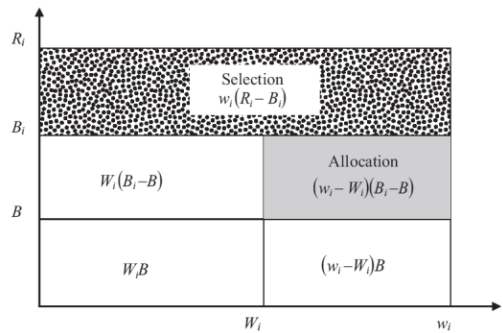

Exhibit 4 – Diagram of the Brinson Model - Bacon and Wright (2019)

The Brinson Model is regarded to be especially useful in analyzing portfolios comprised of equities, derivatives instruments, funds (ETFs and mutual funds), currencies but insufficient to analyze performance attribution on a fixed-income portfolio because "the allocation decision in equity models does not explicitly account for yield-curve positioning (duration) set by fixed-income managers", as explained by McLaren (2002). Exhibit 5 illustrates a typical fixed-income attribution model, adapted from Dai and Elliot (2014/2015). We can see that other effects are in play like carry, yield curve, and spreads.

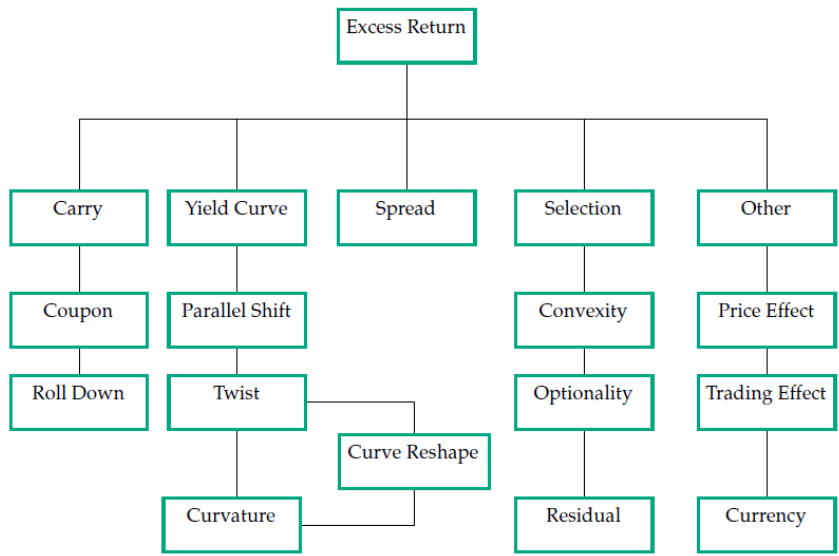

Exhibit 5 – Fixed-Income Attribution Effects

Extensive literature explores other features like multi-period analysis, Carino (1999), geometric attribution, Bacon (2002), and multicurrency attribution, Allen (1991).

Performance Presentation Standards – The CFA Institute sponsored the creation of a presentation standard named Global Investment Performance Standards (GIPS®). As per Bacon (2023), "GIPS provides an ethical framework for the calculation and presentation of the performance history of asset management firms. It is therefore a voluntary standard based on the principles of full disclosure and fair representation. The GIPS standards consist of requirements, which must be followed in order for a firm to claim compliance, and recommendations, which are optional but should be followed because they represent good practice in performance presentation."

AI Agents - The pursue for Artificial General Intelligence (AGI) or the creation of AI systems/agents capable of matching or surpassing human intelligence, has received a significant and relevant push with the release of recent large language models like OpenAI's GPT-4. In a conceptual level, an agent can be characterized by three main parts: brain, perception, and action – as illustrate in Exhibit 6 from Zhiheng et al. (2023). In our context, the LLM acts as the agent's brain, or the controller and is complemented by several components: memory, planning/reasoning, and tool-using.

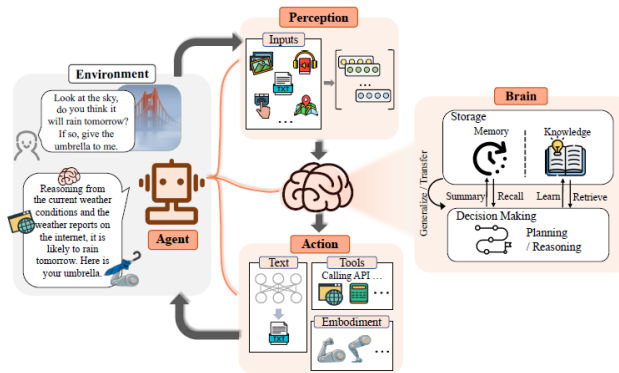

Exhibit 6 – LLM-based agent framework: Brain, Perception, Action - Zhiheng et al. (2023)

On planning or reasoning, the agent breaks down tasks into smaller subtasks, or subgoals, thus enhancing model performance on complex tasks. A prompt technique named Chain of Thought (CoT) has become standard in dealing with complex tasks, Wei et al. (2022). The LLM is explicitly instructed to "think step by step", adding thus a series of intermediate reasoning steps that significantly improves the model's ability to deal with complex reasoning. Other strategies have also been showed to enhance the performance of LLMs like Tree of Thought (ToT) as in Yao et al. (2023), self-consistency as in Wang et al (2023) and ReAct as in Yao et al (2023).

Memory in our context relates to the way the LLM acquires, stores, and retrieves information. Making a parallel to the human brain, there are two main types of memory: short-term and long-term. Short-term memory could be viewed per Weng (2023) as in-context learning, or the model's ability to learn from just a few examples, see Dong et al. (2023) for a survey on the topic. This is a finite memory and restricted by the model's context window length. Long-term memory relates to the capability of retaining and recalling information over longer periods of time, akin to assessing data in an external vector store database, see Milvus (2023).

As for the use of tools, the agent have access to external tools that extend the model capabilities beyond what it was trained on. HuggingGPT developed by Shen et al. (2023) is a framework to handle complicated AI tasks, using ChatGPT to connect various AI models for the resolution of tasks. Gorilla developed by Patil et al. (2023) is a finetuned LLaMA-based model that surpasses the performance of GPT-4 on writing API calls. OpenAI API's function calling and the various plug-ins available in ChatGPT are examples of tools that augment the LLM capability.

LangChain, an LLM development framework by Chase (2022) created several ready-to-use agents applied to particular use cases, like the *create_csv_agent* that interacts with files in tabular format. Under the hood, this agent calls another agent *Pandas DataFrame*, which in turn calls the *Python agent*, which executes LLM generated Python code. The Python agent uses Python REPL (Read-Eval-Print Loop) tool to execute commands provided by LLM. When an agent is executed in LangChain, an "agent executor" chain is initialized. As per LangChain documentation, "the agent executor is the runtime for an agent. This is what actually calls the agent and executes the actions it chooses. Pseudocode for this runtime is below:

```python
next_action = agent.get_action(...)
while next_action != AgentFinish:
    observation = run(next_action)
    next_action = agent.get_action(..., next_action, observation)
return next_action
```

While this may seem simple, there are several complexities this runtime handles, including:

- Handling cases where the agent selects a non-existent tool
- Handling cases where the tool makes an errors
- Handling cases where the agent produces output that cannot be parsed into a tool invocation."

As seen from the pseudocode above, the agent executor follows a similar framework described in Exhibit 6. It comprises **thoughts** that are driven by the LLM, **actions** derived by the thoughts which are commands in Python to be executed and **observations** which are passed back to the LLM for further action. This is done in a loop until the LLM predicts the correct answer has been provided[9].

Evaluation – in order to capture semantic similarity between texts, two scoring methods are useful: ROUGE and Embedding Similarity with Cosine Distance[10].

Lin (2004) first introduced the ROUGE (Recall-Oriented Understudy for Gisting Evaluation) score. This score is typically used to evaluate the quality of text summaries, comparing generated summaries against a reference or ground truth summary, and measures the overlap of n-grams, word sequences, and word pairs between the compared texts. The most commonly used ROUGE metrics are:

---

[9] During the writing of this research, LangChain decided to move the create_csv_agent (and all agents that execute arbitrary SQL and Python code) from its core API package to a secondary API called langchain-experimental, to account for security vulnerabilities linked to arbitrary code executions.

[10] Explanations were adapted from ChatGPT.

- ROUGE-1: measures the overlap of unigrams (single words) between the generated text and the reference text. It essentially counts how many words are common between the two texts, considering both the order and frequency. This metric is useful for assessing the presence of keywords and the basic content overlap.
- ROUGE-2: evaluates the overlap of bigrams (pairs of consecutive words), which helps in understanding how well the generated text captures word pair sequences from the reference text. It's more stringent than ROUGE-1 since it requires word pairs to match, thus being more sensitive to the fluency and structure of the text.
- ROUGE-L: is based on the longest common subsequence (LCS) between the generated text and the reference text. The LCS does not require consecutive words to match but rather looks at the longest sequence of words that appear in both texts in the same order. This metric is particularly good at evaluating the overall fluency and structure of the text as it considers sentence-level structure and word order.

Embedding Similarity with Cosine Distance is another method to capture the semantic similarity meaning of words and phrases. It is also helpful because cosine similarity focuses on the orientation, not the magnitude, of the embedding vectors i.e., it compares the direction of the vectors and not their length. This is important because the length of the vector is usually influenced by the length of the text, but the direction often better represents the content or meaning. Another attractive attribute is that it can handle variability in text effectively. For instance, paraphrased sentences might use different words but have similar meanings. Cosine similarity of embeddings can recognize this similarity, even if the exact words aren't shared. In addition, it is computational and cost efficient in that once embeddings are computed, calculating cosine similarity is computationally trivial, making it suitable for real-time applications.

Risk Measures – in order to measure portfolios with different returns and risk profiles, composite risk measures should be considered within a performance attribution appraisal process. As pointed out by Bacon (2023), "attempting to interpret a portfolio's performance without considering its risk is a misguided venture". Although not explored in this research paper, the relationship between risk and return is fundamental to the investment appraisal process. Measures like Volatility, Sharpe Ratio, and Tracking Error are just some of the measures that capture risk in a single number.

3. Research Question / Hypothesis

The research objective is to use AI agents that can perform the following two main tasks of a typical performance attribution process, when applied to a portfolio comprised of equities:

1: Given a structured data input like a report in a tabular format, the AI agent must explain, interpret, and analyze correctly and accurately the main drivers of excess return, with minimum error. The dataset/report shall be provided as input, based off actual portfolio and benchmark holdings, and returns. This implies that the main components that decomposed the excess return such as allocation effect and selection effect are provided. Analysis can be performed over one or several time periods. A typical use case for the AI agent to be deployed represents the process by which an analyst needs to analyze the performance of one or several funds at the end of the quarter.

2: Calculate the main components of a performance attribution analysis, such as contribution to return, allocation, selection and total effects based on four variables that are given in a structured data input like in a tabular format: returns and weights of a portfolio and its benchmark. Calculation shall be performed at any aggregation level to account for groupings such as by GICS Sectors and Industries, Regions and Countries, Fund Sponsors and Fund Managers, etc. This is often referenced as multi-level analysis where investment decisions are also taken at the parent level. For example, in multi-asset allocation portfolios where the first decision relates to allocating to certain asset classes, like equity vs fixed income – this is the strategic asset allocation decision phase. At a second level, within each asset class, investment decisions in allocating to certain funds, like small-cap vs large-cap, corporate bonds investment grade vs high yield bonds – this is the tactical asset allocation decision phase. In this case, the AI agent will implement calculations using arithmetic excess returns over one time period at different levels. A variant research question is to pose questions based on the same table adapted from actual CIPM (Certificate in Investment Performance Measurement, managed by the CFA Institute) preparatory exams A typical use case for the AI agent to be deployed represents the process by which an analyst needs to quantify the excess return decomposition without the use of an attribution system.

On objective 1, the intention is to evaluate whether an AI system can replicate a typical performance appraisal process in which calculations are performed by a software (results are assumed to be correct and accurate), and interpretation combined with in-depth analysis are performed on top of these results. Evaluation is performed on descriptive texts that explain the decomposition of total contribution, especially through the combination of factors such as overweight/underweight, outperformance/underperformance, and on the numerical evidence produced by the model. On 2, the intention is to evaluate whether an AI agent can correctly perform mathematical calculations and produce correct results be it in the form of an explicit request to calculate a complete attribution report or in the form of questions answers multiple choices (QAMC) tests.

The hypothesis is thus that despite LLMs being prone to hallucination, an AI agent running in a suitable foundation model with practical prompt engineering and RAG process, can produce a performance attribution analysis and be able to calculate its main components (Allocation, Selection and Total Effects) with minimum error. One important premise underlying the hypothesis and the research questions is that the AI Agent is used "as is", meaning that there is no change to its original code, despite being an open-source project. This decision was taken in order to create a baseline for future research projects and also to save research time.

In practical terms, a standard performance attribution dataset for long-only equity portfolio might contains columns like: Breakdown by GICS Sector or Currency, Portfolio and Benchmark Weights, Portfolio and Benchmark Returns, Variation in Weight, Variation in Return, Allocation Effect, Selection Effect and Total Contribution.

As illustration, consider the performance attribution output in Exhibit 7 which depicts the decomposition of the excess return of an equity portfolio versus its benchmark calculated using the Brinson-Fachler Model. It contains columns like: Breakdown by Group (GICS Sector or Country), Portfolio and Benchmark Weights, Portfolio and Benchmark Returns, Variation in Weight, Variation in Return, Allocation Effect, Selection Effect and Total Contribution. In this case, grouping is done by GICS sector, hence there are twelve rows plus aggregation.

Although the use of structured data as an external knowledge layer to LLMs and the subsequent querying/instruction is somewhat simple to implement thanks to applications like LangChain, a performance attribution table presents a different set of challenge because it involves combined analysis across at least six different columns for an equity portfolio (industry sector, portfolio weight, benchmark weight, portfolio return, benchmark return, allocation effect, selection effect and total contribution).

| GICS Sector | Portfolio Weight | Benchmark Weight | Portfolio Return | Benchmark Return | Variation in Weight | Variation in Return | Allocation Effect | Selection Effect | Total Contribution |
|---|---|---|---|---|---|---|---|---|---|
| Consumer Staples | 14.5% | 8.8% | 11.3% | 9.7% | 5.7% | 1.6% | 0.15% | 0.23% | 0.38% |
| Utilities | 13.8% | 7.2% | 8.9% | 6.7% | 6.6% | 2.2% | -0.03% | 0.30% | 0.27% |
| Consumer Discret. | 8.9% | 9.3% | 9.4% | 7.3% | -0.4% | 2.1% | 0.00% | 0.19% | 0.19% |
| IT | 8.6% | 11.2% | 10.6% | 8.2% | -2.6% | 2.4% | -0.03% | 0.21% | 0.18% |
| Real Estate | 10.3% | 6.7% | 9.4% | 7.9% | 3.6% | 1.5% | 0.03% | 0.15% | 0.18% |
| Energy | 3.7% | 7.5% | 3.2% | 4.3% | -3.8% | -1.1% | 0.11% | -0.04% | 0.07% |
| Materials | 5.3% | 8.9% | 4.7% | 5.2% | -3.6% | -0.5% | 0.07% | -0.03% | 0.04% |
| Industrials | 6.8% | 9.3% | 5.7% | 6.1% | -2.5% | -0.4% | 0.03% | -0.03% | 0.00% |
| Financials | 7.7% | 13.6% | 4.3% | 5.7% | -5.9% | -1.4% | 0.09% | -0.11% | -0.02% |
| Health Care | 11.6% | 9.4% | 7.1% | 7.7% | 2.2% | -0.6% | 0.01% | -0.07% | -0.06% |
| Communication | 6.8% | 8.1% | 9.3% | 10.3% | -1.3% | -1.0% | -0.04% | -0.07% | -0.11% |
| Cash | 2.0% | 0.0% | 0.4% | 0.0% | 2.0% | 0.4% | -0.14% | 0.00% | -0.13% |
| Total | 100.0% | 100.0% | 8.1% | 7.2% | 0.0% | 1.0% | 0.2% | 0.8% | 1.0% |

Exhibit 7: Performance Attribution Report of an Equity Portfolio by GICS Sector

Although not part of this research, for a fixed income portfolio performance analysis, there are even more columns to be analyzed, as depicted in Exhibit 3. Apart from Allocation and Selection Effects, at least two other effects are required: Shift and Twist Effects, related to the Yield Curve. However, the framework can be extended to as much as eight to ten additional columns.

As an integral and important part of the investment decision process, performance attribution analysis provides value in that it allows for independent feedback into the investment process and gives transparency to various internal stakeholders and external clients on the portfolio sources of returns and risks. One advantage of using an agent AI in this context would be its ability to analyze several datasets in one go, thus reducing the cognitive burden of a human analysis and decreasing costly content revision and time to publishing/communication. In addition, an agent might also be used for other purposes such as generating insights or detecting patterns or anomalies.

## 4. Methodology

4.1 Objective 1 Methodology

The following methodologies are used to accomplish objective #1:

1: Performance Attribution: several reports were created that mirror an equity fund offering from a typical asset manager: defensive fund with low variation in weights between the portfolio and the benchmark; growth-oriented funds with higher exposure to growth sectors like information technology; and value-oriented with higher exposure to sectors like utilities and healthcare. For each of these funds, four quarterly reports covering one calendar year resulting in a total of twelve distinct reports were generated. Results are grouped by GICS sectors, and each report is indexed by their respective portfolio name, benchmark name and date. All reports obey the same format: one header, twelve rows (one for each GICS sector plus Cash), thirteen columns (GICS Sector, Portfolio and Benchmark Weights, Portfolio and Benchmark Returns, Variation in Weight, Variation in Return, Allocation Effect, Selection Effect, Total Contribution, Period, Fund and Benchmark Name). A sample report is available in the Appendix.

From Exhibits 1, 2 and 3, several keywords which are part of a typical attribution analysis write up can be identified, such as overweight/underweight, outperformance/underperformance or neutral, like in the following sentence: "the Energy sector has a positive allocation effect of 0.35%. This was due to the portfolio being 1.5% overweight in this sector compared to the benchmark. The overweight in a sector that outperformed the overall benchmark return led to this positive result." Combining these keywords or factors with numerical evidence in a correct way is the essence of Objective 1 and ultimately determines the LLM's accuracy in explaining the reasons for that particular effect to be positive, negative or zero.

2: AI Agent: LangChain provides different agents function but because the performance report is formatted as tabular and consequently saved as a *csv* file, we use the *create_csv_agent* agent function that answers questions about a CSV file - see Appendix for a complete description of the *create_csv_agent* function. The main inputs are:

   a. LLM model: model name and temperature. For model, GPT-4 was chosen because it is still state-of-the-art in all benchmarks related to reasoning and instruction. For temperature, which is a parameter that controls the level of randomness on the response, we used 0 temperature which is very deterministic.
   b. Path to the CSV file
   c. Agent Type: Different agent types are available, and we used the *OpenAI Functions Agent.* This was chosen because the GPT-4 model was explicitly fine-tuned to detect when a function should be called and to respond with the inputs that should be passed to another function. GPT-3.5 model was also fine-tuned on instruction, but we found that it is not able to run some of the more complex prompts.

3: Prompt: to assess whether LLMs could reason and be able to explain the drivers of excess return and to evaluate the responses afterwards, prompts were explicitly engineered with: i) specific input instructions, using few-shot and zero-shot in-context learnings, and with: ii) specific output instructions.

The input instructions contains either one or two pieces of in-context learning (ICL) for the few-shot prompts and no ICL for the zero-shot prompt. In total, four prompts are tested: three few-shot and one zero-shot. One specific ICL is common among all few-shot prompts, and it refers to a direct request for the LLM to answer about the allocation and selection effects in a format that forces the model to choose a combination of keyworks/factors that explains that particular effect, thus forcing the model to reason. Exhibit 8 depicts the prompt components for the prompt named "*few_shot_1*".

Exhibit 8: Few Shot Prompt – model 1

Two additional few-shot prompts are constructed with additional learnings in order to test for improved accuracy, namely "*few_shot_2*" and "*few_shot_3*". On the "*few_shot_2*" prompt, we explicitly explain the reasoning behind positive and negative allocation effect. On the "*few_shot_3*" prompt, we provide numerical examples explaining the reasoning behind positive and negative allocation effect. It was decided to only add ICL regarding the allocation effect because it is the most challenging effect to explain as it requires a combination of two keywords/factors. Exhibits 9 and 10 display these additional ICL.

```
Attention: These additional explanations helps with your reasoning, especially in the allocation effect:
Positive Allocation occurs in two instances: 1) when the portfolio is 'overweigthed' versus the benchmark and the benchmark return 'outperform's the benchmark total return
or 2) when the portfolio is 'underweighted' versus the benchmark and the benchmark return 'underperforms' the benchmark total return.
Negative allocation occurs in two instances: 1) when the portfolio is 'overweigthed' versus the benchmark and the benchmark return 'underperforms' the benchmark total
return or 2) when the portfolio is 'underweighted' versus the benchmark and the benchmark return 'outperforms' the benchmark total return.
```

Exhibit 9: Few Shot Prompt – model 2

```
- The 'Consumer Discret.' sector had a 'positive' allocation effect of '0.03'. This was due to the fund being 'overweight' in this sector compared to the benchmark and due
to the the benchmark total return of '0.04' 'outperformed'.
- The 'Consumer Discret.' sector had a 'positive' allocation effect of '0.03'. This was due to the fund being 'underweight' in this sector compared to the benchmark and due
to the the benchmark total return of '0.06' 'underperformed'.
- The 'Consumer Discret.' sector had a 'negative' allocation effect of '-0.03'. This was due to the fund being 'overweight' in this sector compared to the benchmark and due
to the the benchmark total return of '0.06' 'underperformed'.
- The 'Consumer Discret.' sector had a 'negative' allocation effect of '-0.03'. This was due to the fund being 'underweight' in this sector compared to the benchmark and du
to the the benchmark total return of '0.04' 'outperformed'.
```

Exhibit 10: Few Shot Prompt – model 3

The zero-shot prompt contains the same output instruction and is displayed in Exhibit 11.

```python
def construct_prompt_zeroshot_2(sector):

    prompt = f"""
    Analyze performance attribution for the {sector} sector using the Brinson model.
    Provide:
    1. Two bullet points: one for allocation and one for selection, with reasoning and numerical evidence.
    2. Convert the bullet points in a CSV format with columns: Sector, Effect Type, Value, Sector Weight, Sector Performance.
    3. CSV column "Sector Weight" choices: Underweight or Overweight. On selection, write None.
    4. CSV column "Sector Performance" choices: Outperformance or Underperformance.
    - Output should contain only bullet points and the CSV format, separate by 'CSV Format:'.
    - Write maximum 3 sentences per bullet point.

    """
    return prompt
```

Exhibit 11: Zero Shot Prompt

The output request contains instructions for the LLM to generate two kinds of responses, which are later used for evaluation: i) descriptive text as two bullet points explaining the allocation and selection effects, and ii) conversion of the descriptive text to a CSV format containing the relevant keywords/factors of the text previously generated – see Exhibit 12. Here we rely on the LLM's intrinsic capacity to correctly identify entities and then perform the conversion with little effort. The descriptive text output should contain two or three sentences per bullet point/effect, and it is afterwards converted to a csv file for evaluation. The second output containing the CSV part has three rows, one for the header (containing columns Sector, Effect Type, Value, Sector Weight, Sector Performance) and two rows containing the factors with numerical values for each effect – Exhibit 12 displays a sample generated output. A post processing script saves these two outputs in separated csv files, which are later used for evaluation.

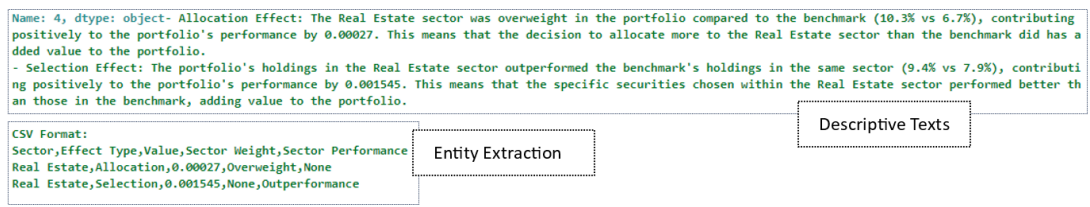

Exhibit 12: Sample generated output

Given the performance attribution report is broken down by GICS sectors, the AI agent executes one API call per sector through a loop executed externally to the LLM. An alternative is to execute one API/agent call per fund (or chunk), implying a loop to be requested in the prompt – this was not tested, and it is left for future work.

4: Evaluation: from the two outputs described previously, descriptive texts in the form of bullet points and a CSV formatted output, two distinctive evaluation methods are designed.

Method 1: we use semantic similarity between the descriptive texts and hard-coded sentences, used as baseline. Two methods are employed: ROUGE and embedding similarity using cosine distance. On ROUGE, we calculate the F1 score which combines precision and recall into a single metric that balances both. On embedding similarity using cosine distance, a text encoding model called *all-mpnet-base-v2* developed by HuggingFace and OpenAI's *text-embedding-ada-002* are used to create the embeddings. Comparison is made using the cosine similarity metric, which varies from 0 to 1 for each text. The maximum score obtained is then 2 out of 2 per sector (1 for each effect) and 24 out of 24 per portfolio (2 for each sector).

The hard-coded sentences were built using simple instructions derived from comparing weights and returns, following the format

Method 2: we compare the factor values of each effect for each sector (columns Value, Sector Weight, Sector Performance) in the CSV files with their trues values, which are appended to the CSV file after the entity extraction stage is done. Similarity is measured by accuracy (fraction of correct predictions) where the maximum value that can be obtained is 6 out of 6 per sector (3 for each effect) and 72 out of 72 per portfolio (6 for each sector).

5: Algorithm: the pseudocode to generate the outputs from a performance attribution analysis report is illustrated in Exhibit 13.

```
Input:    A performance attribution analysis report R
          LLM model L with Temperature T
          LangChain Agent A
          Agent Type Function AT
          A prompt template Pi belonging to a collection of prompts P
          An embedding model E

Output:   A response from LLM Re
          A csv file containing descriptive texts DT
          A csv file containing factors F
          A csv file containing hard-coded sentences for evaluation HC
          A csv file containing cosine similarity scores Co
          A csv file containing factor hit ratio scores Fh
          Evaluation score Ev

//Querying LLM and generating output
Read and Convert input file R with n sectors into a pandas dataframe df
Initialize Agent A with Model L and Agent Type AT
for all i ≤ n do
          Capture true values from df
          Generate hard-coded sentences for evaluation based on true values from df
          Execute Agent A using prompt P
          for all p in P do
                    Extract from response Re:
                              Descriptive Texts
                              CSV factors
                    Create output file DT and append new fields
                    Create output file F and append new fields
          end for
end for
//Evaluation of output files
//First method: semantic similarity
Read and Convert output file DT with n sectors into a pandas dataframe df_1
Read and Convert output file HC with n sectors into a pandas dataframe df_2
Convert df_1 and df_2 into embeddings using model E
for all i ≤ n do
          Read df_1 for i
          Read df_2 for i
          Measure cosine similarity
          Create output file Co and append new fields
end for
//Second method: factors hit ratio
Read and Convert output file F with n sectors and m columns into a pandas dataframe df_3
for all i ≤ n do
          for all j ≤ m do
                    Compare column j and column j+1
                    Calculate score
                    Create output file Fh and append new fields
          end for
end for
```

Exhibit 13: Objective #1 algorithm

4.2 Objective 2 Methodology.

The following methodologies are used to accomplish objective #2:

1: Performance Attribution: the same reports used previously in the making of objective #1 are employed in this objective but only with four variables: fund's and its benchmark's returns and weights. This objective requires the calculation of a complete attribution analysis report where multiple levels are also present. For each of the three funds, four reports with one aggregation (GICS Sectors and parent) resulting in a total of twelve distinct reports were used. All reports obey the same format: one header, multiple rows, and six columns (Group Parent, Group, Portfolio and Benchmark Weights, Portfolio and Benchmark Returns). A sample report is available in the Appendix. The AI agent will implement calculations using arithmetic excess returns over one time period, thus resulting in 12 tables to be evaluated.

As in objective #1, the Brinson-Fachler model is used. Two additional sets of calculations are performed: two types of multiple level attribution, which are explained below.

1.1: Multiple Level attribution

Multiple level performance attribution analysis uses the same principles and equation as the single level but there is a choice of which level the calculations are performed. In our case where there are two levels, two sets of attribution analysis are performed, using the same set of weights and returns for each level – macro and micro attributions. Example below illustrate these concepts - extracted from Bacon and Wright (2019).

Consider a strategic asset allocation policy where decisions are made between investing in Value or Growth portfolio managers – Exhibit 14 illustrates the portfolio and benchmark weights and returns (in % as depicted) from Bacon and Wright (2019). The fund managers then decide where to allocation within their area of expertise. At the **macro level**, investment decision is to overweight Value managers by 3% (78 − 75%) and to underweight Growth managers by the same amount. At the **micro level**, investment decision is to overweight large-cap value while underweight small-cap value and large-cap growth. These decisions paid off, for the arithmetic excess return is positive 98bps.

| | Fund Weight | Fund Return | Benchmark Weight | Benchmark Return | |
|---|---|---|---|---|---|
| **Total** | 100% | 0.95 | 100% | −0.03 | Macro level decisions |
| Value Portfolio Manager | 78% | 0.99 | 75% | 0.32 | |
| *Small-cap value equities* | 20% | 2.39 | 25% | 1.52 | Micro level decisions |
| *Large-cap value equities* | 58% | 0.51 | 50% | −0.28 | |
| Growth Portfolio Manager | 22% | 0.82 | 25% | −1.08 | Macro level decisions |
| *Large-cap growth equities* | 22% | 0.82 | 25% | −1.08 | |

Exhibit 14 – Multiple Level performance attribution - Bacon and Wright (2019)

Starting at the lower segment and using equations 2.3 to 2.6, the allocation and selection effects for the small-cap equities are:

$$Allocation = (20\% - 25\%) \times [1.52\% - (-0.03\%)] = -0.08\%$$

$$Selection = 20\% \times (2.39\% - 1.52\%) = 0.17\%$$

Using the same formulas for the two other segments, we arrive at the results displayed in Exhibit 15 from Bacon and Wright (2019). Interpretation is that the main cause of excess return is due to security selection. On allocation, the Value Manager detracted value due to their decision to underweight small cap in favor of large cap while the Growth Manager added value.

| Return attribution (segment level) | Fund Weight | Selection + Interaction | Allocation | Total |
| --- | --- | --- | --- | --- |
| Total | 100% | 1.05 | −0.07 | 0.98 |
| Value Portfolio Manager | 78% | 0.63 | −0.10 | 0.53 |
| *Small-cap value equities* | 20% | 0.17 | −0.08 | 0.10 |
| *Large-cap value equities* | 58% | 0.46 | −0.02 | 0.44 |
| Growth Portfolio Manager | 22% | 0.42 | 0.03 | 0.45 |
| *Large-cap growth equities* | 22% | 0.42 | 0.03 | 0.45 |

Exhibit 15 – Segment level performance attribution - Bacon and Wright (2019)

At the parent or macro level, a second return attribution analysis calculates the return attribution at the portfolio manager level, using the same formulas as before. For example, for the Value Portfolio Manager, the effects are as follows:

$$Allocation = (78\% - 75\%) \times [0.32\% - (-0.03\%)] = 0.01\%$$

$$Selection = 78\% \times (0.99\% - 0.32\%) = 0.52\%$$

Note that benchmark segment return of 0.32% is calculated as the weighted average return of the Value Portfolio Manager sub-group:

$$0.32\% = \frac{(25\% \times 1.52\%) + (50\% \times -0.28\%)}{25\% + 50\%} \tag{4.1}$$

Using the same formulas for the two other segments, we arrive at the results displayed in Exhibit 16 from Bacon and Wright (2019). The interpretation is that the main cause of excess return is still due to security selection, but allocation is now positive, because the evaluation is done at a different allocation decision level. The asset allocator decision to overweight value versus growth was additive to performance.

| Return Attribution (Portfolio Manager level) | Selection + Interaction | Allocation | Total |
|---|---|---|---|
| Total | 0.94 | 0.04 | 0.98 |
| **Value Portfolio Manager** | **0.52** | **0.01** | **0.53** |
| Small-cap value equities | – | – | – |
| Large-cap value equities | – | – | – |
| **Growth Portfolio Manager** | **0.42** | **0.03** | **0.45** |
| Large-cap growth equities | – | – | – |

Exhibit 16 – Portfolio Manager Level performance attribution - Bacon and Wright (2019)

As seen before, the formulas and equations used on a multiple level attribution analysis are the same, and such a report can consequently carry any number of aggregation dependencies. What is relevant is to decide which level the calculations are performed so that the intermediate calculations like in equation 4.1 are correctly performed. In this example, the benchmark sector return at the fund manager is given but in our input tables that is not going to be the case.

1.2: Question-Answer (QA)

Two types of QA were designed, mimicking questions from actual exams: i) question-answer-multiple-choice (QAMC), ii) question to calculate (QCalc). On QAMC, a set of four choices are given and the question is posed similarly to Exhibit 17 – extracted from Bacon and Wright (2019). On QCalc, the question is simply a request to calculate a given effect from the table provided.

The exhibit below shows the weights and returns of a portfolio and its benchmark.

| | Portfolio Weight | Benchmark Weight | Portfolio Return | Benchmark Return |
|---|---|---|---|---|
| Health Care | 10% | 20% | 3% | 2% |
| Utilities | 30% | 30% | 4% | 4% |
| Consumer Goods | 60% | 50% | 7% | 8% |
| **Total** | **100%** | **100%** | **5.7%** | **5.6%** |

According to the Brinson–Fachler attribution model, the allocation effect from the Consumer Goods sector is *closest* to:

A  −0.50%.
B  0.24%.
C  0.80%.

Exhibit 17 – QAMC question - Bacon and Wright (2019)

The essence of this objective thus follows the mechanics of the examples given above. Given a table containing weights and returns, and applying equations 2.3 to 2.6 and 4.1, allocation and selection effects are calculated for each level present in the table.

2: AI Agent: same as used in Objective #1.

3: Prompt: to assess whether LLMs could calculate single or multiple level performance attribution analytics, prompts were explicitly engineered with: i) specific inputs, using few-shot in-context learning, with: ii) specific planning and output instructions.

The input contains explicit explanations in a formulaic fashion on how to calculate the allocation and selection effects and all the intermediate steps, in case of multiple level calculation. The output contains planning instructions on how to execute these formulae in order to calculate intermediate steps and the final results. Two prompts were created: one calculates allocation and selection effects by the lowest segment level (micro attribution level analysis) named *"micro prompt"* and the other calculates at the highest segment level (macro attribution level analysis), named *"macro prompt"*. The *micro prompt* has four descriptive expressions and two output requests, while the *macro prompt* has eight descriptive expressions and step-by-step planning instructions on how to calculate intermediate and final results. Sample prompts are displayed in Exhibit 18.

The output from the micro prompt is formatted as a table containing the allocation, selection, and total contribution by sector and at the top-level – see Exhibit 19. This table is then converted to a pandas dataframe that is used in the evaluation stage. The output from the macro prompt displays the final results at the top-level, and it is also converted to a dataframe for later evaluation – see Exhibit 20.

```
generate_prompt_micro_2 = """
Calculate performance attribution using the Brinson Model for a specific Fund and Period with the provided data. Follow these formulas:
- Allocation Effect = (Portfolio Weight - Benchmark Weight) * (Benchmark Return - Benchmark Total Return)
- Selection Effect = Portfolio Weight * (Portfolio Return - Benchmark Return)
- Total Contribution = Allocation Effect + Selection Effect
- Benchmark Total Return = Weighted average of Benchmark weights and Benchmark returns
```
[Formulae learning]

```
This is a multi-level problem where sectors below to 'GICS Type'.

Objective:
1. Calculate 'Allocation Effect', 'Selection Effect', and 'Total Contribution' for all 'GICS Sector'. Output in CSV format, with the following columns:
'GICS Sector', 'Allocation Effect', 'Selection Effect', 'Total Contribution', 'Fund', 'Period'. Each record should be on a new line.
2. Calculate 'Allocation Effect', 'Selection Effect', and 'Total Contribution' for all 'GICS Type'. Output in CSV format, with the following columns:
'GICS Type', 'Allocation Effect', 'Selection Effect', 'Total Contribution', 'Fund', 'Period'. Each record should be on a new line.
3. Output should contain only the CSV format, initiated by 'CSV Format:'. No additional sentences, separators, wordings.
```
[Output instruction]

```
Attention: Use pandas for calculations. Ensure accuracy and adherence to the formulas.
Attention: Ensure the output contains only the CSV formatted data.

"""

generate_prompt_macro = """
Your objective is to calculate performance attribution based on the Brinson Model for a specific Fund and Period with the provided data. Follow these formulas:
'Allocation Effect' = (Portfolio Weight - Benchmark Weight) * (Benchmark Return - Benchmark Total Return)
'Selection Effect' = Portfolio Weight * (Portfolio Return - Benchmark Return)
'Total Contribution' = 'Allocation Effect' + 'Selection Effect'
'Portfolio Contribution to Return' = Portfolio Weight * Portfolio Return
'Benchmark Contribution to Return' = Benchmark Weight * Benchmark Return
'Benchmark Total Return' is calculated as the weighted average between Benchmark weights and Benchmark returns.
'Portfolio Return' is the sum over 'Portfolio Contribution to Return' divided by the specific level/segment total weight
'Benchmark Return' is the sum over 'Benchmark Contribution to Return' divided by the specific level/segment total weight
```
[Formulae learning]

```
This is a multi-level problem where sectors belong to 'GICS Type'. Think carefully on how to aggregate before giving your answers.
Your goal is to calculate Allocation and Selection effect for all 'GICS Type' from the top-down or highest segment, i.e at the 'GICS Type' level, following these steps:
Step 1: Calculate 'Benchmark Total Return'
Step 2: Calculate the 'Portfolio Return' and 'Benchmark Return' at each 'GICS Type' level.
Step 3: Use these results to calculate the 'Allocation Effect', 'Selection Effect' at each 'GICS Type' level.
Attention: Your output is in a CSV format with the following columns: GICS Type, Effect Type, Value, Fund, Period. Each record should be on a new line.
Attention: 'Value' corresponds to the numerical values calculated in Step 3.
Attention: Output should contain the CSV format, initiated by 'CSV Format:', but no additional sentences.
Attention: Use pandas
Attention: Be intentional, You are expected to provide numerical answers and perform calculations following the formulas and the steps provided.
```
[Output instruction]

```
"""
```

Exhibit 18 - Sample Prompts

```
1. Using the Brinson Model, the Allocation and Selection effect for all 'GICS Sector' are as follows:

| GICS Sector       | Allocation Effect | Selection Effect | Total Contribution |
|-------------------|-------------------|------------------|--------------------|
| Cash              | -0.001430         | 0.000083         | -0.001346          |
| Communication     | -0.000410         | -0.000680        | -0.001090          |
| Consumer Discret. | -0.000006         | 0.001869         | 0.001863           |
| Consumer Staples  | 0.001454          | 0.002320         | 0.003774           |
| Energy            | 0.001082          | -0.000407        | 0.000675           |
| Financials        | 0.000855          | -0.001078        | -0.000223          |
| Health Care       | 0.000121          | -0.000696        | -0.000575          |
| IT                | -0.000273         | 0.002064         | 0.001791           |
| Industrials       | 0.000262          | -0.000272        | -0.000010          |
| Materials         | 0.000701          | -0.000265        | 0.000436           |
| Real Estate       | 0.000271          | 0.001545         | 0.001816           |
| Utilities         | -0.000296         | 0.003036         | 0.002740           |

2. Using the Brinson Model, the Allocation and Selection effect for all 'GICS Type' are as follows:

| GICS Types | Allocation Effect | Selection Effect | Total Contribution |
|------------|-------------------|------------------|--------------------|
| Cyclical   | 0.001400          | 0.003183         | 0.004583           |
| Defensive  | 0.000932          | 0.004336         | 0.005269           |
```

Exhibit 19 - Micro Prompt output

```
The Allocation and Selection effects for each 'GICS Type' are as follows:

For 'Cyclical':
- Allocation Effect: 0.000058
- Selection Effect: 0.004525

For 'Defensive':
- Allocation Effect: 0.000118
- Selection Effect: 0.005151
```

Exhibit 20 - Macro Prompt output

As for the QA method, two types of prompts were designed and they share the same structure: formulae learning, a question about the allocation or selection effects and multiple choices – see Exhibit 21. Question structure is displayed in Exhibit 22.

```
def format_question_2(question):
    """Formats the question for GPT-4."""
    options_text = "\n".join([f"{letter}: {opt}%" for letter, opt in zip("ABCD", question['options'])])

    if "is closest to" in question["question"]:
        # It's a "value" question
        prompt = f"""
Question below asks you to calculate performance attribution based on the Brinson Model.

You know that for a specific Fund and Period:
'Allocation Effect' = (Portfolio Weight - Benchmark Weight) * (Benchmark Return - Benchmark Total Return)
'Selection Effect' = Portfolio Weight * (Portfolio Return - Benchmark Return)
'Total Contribution' = 'Allocation Effect' + 'Selection Effect'
'Benchmark Total Return' is calculated as the weighted average between columns 'Benchmark weight' and 'Benchmark return'

You are giving four choices to answer the question. You must choose one of the answers and your output is just the first letter of that answer, nothing else.
Answer 'E' if you do not know the answer.
Attention: Data provided contains various funds and periods. Choose correctly.
Question:
Solve step-by-step, first extract the Fund and Period provided in the question, second calculate 'Benchmark Total Return' for all 'GICS Sector' filtered by the Fund and Period extracted,
third calculate Allocation and Selection Effect for all 'GICS Sector' filtered by the Fund and Period extracted; and finally answer the following: {question["question"]}
Choices:
{options_text}
Answer:"""
    else:
        # It's a "calculation" question
        prompt = f"""
Question below asks you to calculate performance attribution based on the Brinson Model. Make sure to output the answer as a numerical value rounded to five decimal places.

You know that for a specific Fund and Period:
'Allocation Effect' = (Portfolio Weight - Benchmark Weight) * (Benchmark Return - Benchmark Total Return)
'Selection Effect' = Portfolio Weight * (Portfolio Return - Benchmark Return)
'Total Contribution' = 'Allocation Effect' + 'Selection Effect'
'Benchmark Total Return' is calculated as the weighted average between columns 'Benchmark weight' and 'Benchmark return'

Note: Your answer should be a single numerical value rounded to five decimal places.
Attention: Data provided contains various funds and periods. Choose correctly.
Question:
Solve step-by-step, first extract the Fund and Period provided in the question, second calculate 'Benchmark Total Return' for all 'GICS Sector' filtered by the Fund and Period extracted,
third calculate Allocation and Selection Effect for all 'GICS Sector' filtered by the Fund and Period extracted; and finally answer the following: {question["question"]}
Note: Your answer should be a single numerical value rounded to five decimal places, nothing else.
Answer:"""
    return prompt
```

Exhibit 21 - QAMC and QCalc prompts

```
# Determine the question type based on the iteration
if i < num_questions // 2:
    question_type = "value"
    question_text = f"The {column_name.replace('_', ' ').title()} from the {row['GICS Sector']} sector, in fund {fund}, in the period {period}, is closest to:"
else:
    question_type = "calculation"
    question_text = f"Calculate the {column_name.replace('_', ' ').title()} from the {row['GICS Sector']} sector, in fund {fund}, in the period {period}:"
```

Exhibit 22 - Question types

4: Evaluation: from the two methods described previously, two distinctive evaluation methods are designed.

Method 1: we compare the allocation, selection, and total effects values at each aggregation level (macro and micro levels) for each grouping, and for each sector with their trues values. In the micro level calculation, in the case of a twelve GICS sector report grouped in a parent with three different types (Cyclical, Defensive, Sensitive), there are twelve "sector values" plus three "type values" per fund per period, for a total of 60 values per fund. In the macro level, there are three "type values" per fund per period, for a total of twelve values per fund.

Method 2: we generated 140 questions, half of them with multiple choice (QAMC) and the other half with questions to calculate an effect of a specific sector (QCalc) – for a total of seventy maximum points per question type.

5: Algorithm: the pseudocode to generate the outputs for the micro and macro calculations, and for the QAMC is illustrated in Exhibit 23. QCalc pseudocode is very similar to QAMC and it is not displayed.

| Input: | A report $R$ comprised of returns and weights data (no attribution effects)<br>LLM model L with Temperature $T$<br>LangChain Agent $A$<br>Agent Type Function $AT$<br>A prompt template $Pi$ belonging to a collection of prompts $P$ | Input: | A performance attribution analysis report $R$<br>LLM model L with Temperature $T$<br>LangChain Agent $A$<br>Agent Type Function $AT$<br>A prompt template $Pi$ belonging to a collection of prompts $P$ |
|---|---|---|---|
| Output: | A response from LLM $Re$<br>A csv file containing attribution calculations at the lowest segment $Al$<br>A csv file containing attribution calculations at the highest segment $AL$<br>A csv file containing ground-truth values for evaluation $HC$<br>A csv file containing accuracy ratio scores $Fh$<br>Evaluation score $Ev$ | Output: | A response from LLM $Re$<br>A csv file containing answers either a value (letter A, B, C, D or E) or a calculated number $A$<br>A csv file containing ground-truth values for evaluation $HC$<br>A csv file containing accuracy ratio scores $Fh$<br>Evaluation score $Ev$ |

```
//Querying LLM and generating output
Read and Convert input file R with n sectors into a pandas dataframe df
Initialize Agent A with Model L and Agent Type AT
for all i ≤ n do
        Capture ground-truth values from df
        Create output HC
        Execute Agent A using prompt P
        for all p in P do
                Extract from response Re:
                        Al and AL
                Create output file Al and append new fields
                Create output file AL and append new fields
        end for
end for
//Evaluation of output files
//Method: output accuracy
Read and Convert output file Al and AL with n sectors and m columns into a pandas dataframe df_3
for all i ≤ n do
        for all j ≤ m do
                Compare column j and column j+1
                Calculate score
                Create output file Fh and append new fields
        end for
end for
```

```
//Querying LLM and generating output
Read and Convert input file R with n sectors into a pandas dataframe df
Initialize Agent A with Model L and Agent Type AT
for all i ≤ n do
        Generate random question from report R
        Generate answers (one true and three false)
        Capture ground-truth values from df
        Create output HC
        Execute Agent A using prompt P
        for all p in P do
                Extract from response Re:
                        Letter or Calculated Value
                Create output file A and append new fields
        end for
end for
//Evaluation of output files
//Method: output accuracy
Read and Convert output file A into a pandas dataframe df_3
for all i ≤ n do
        for all j ≤ m do
                Compare column j and column j+1
                Calculate score
                Create output file Fh and append new fields
        end for
end for
```

Exhibit 23 – Objective #2 algorithms

## 5. Results

In this section, results are presented and discussed.

Objective #1:

The output from the four types of prompts used (*few_shot_1, few_shot_2, few_shot_3, zero_shot*) were measured using semantic similarity and an accuracy ratio. In general, GPT4 did not experience any issues related to inference time, producing a text or numerical results, and following output instruction, while running the prompts sector by sector, fund by fund, in sequence.

Method 1: Semantic Similarity

Using the cosine similarity metric, the results obtained with the OpenAI embedding model are all over 94% while using the sentence transformers embedding, it ranges from 81 to 97%. Comparing the prompt types, there are no big differences between the various *few_shot* types, regardless of the embedded model used. Not surprisingly, zero-shot underperforms in relation to few-shot, in both embedded models. In terms of differences between allocation and selection sentences, no big discrepancies are observed in OpenAI's embeddings, but differences are more pronounced in the ST's embeddings. Exhibit 24 summarizes these findings.

|  | **OpenAI** | | **Sentence Transformers** | |
| --- | --- | --- | --- | --- |
|  | **Allocation** | **Selection** | **Allocation** | **Selection** |
| **few_1** | 99% | 98% | 97% | 92% |
| **few_2** | 99% | 98% | 97% | 92% |
| **few_3** | 98% | 97% | 94% | 89% |
| ***Few Shot*** | **99%** | **97%** | **96%** | **91%** |
| **zero_1** | 95% | 94% | 83% | 81% |
| ***Zero Shot*** | **95%** | **94%** | **83%** | **81%** |

OpenAI Embeddings: *text-embedding-ada-002*
Sentence Transformers: *all-mpnet-base-v2*

Exhibit 24 – Semantic similarity by cosine distance

As for the ROUGE scores, using the F-measure or F1 score, most of the results are similar to the previous method. Across all score types, comparing the prompt types, there is no big differences between the various types. Zero-shot underperforms in relation to few-shot across all score types. A different conclusion is that selection sentences score higher than allocation ones, especially on the ROUGE-L, indicating that reasoning worked better in the selection sentences. Exhibit 25 summarizes these findings.

|  | Allocation | | | Selection | | |
| --- | --- | --- | --- | --- | --- | --- |
|  | ROUGE-1 | ROUGE-2 | ROUGE-L | ROUGE-1 | ROUGE-2 | ROUGE-L |
| few_1 | 73% | 52% | 65% | 70% | 53% | 70% |
| few_2 | 72% | 51% | 65% | 69% | 51% | 68% |
| few_3 | 72% | 51% | 65% | 69% | 51% | 69% |
| *Few Shot* | **72%** | **51%** | **65%** | **69%** | **52%** | **69%** |
| zero_1 | 50% | 18% | 31% | 41% | 20% | 33% |
| *Zero Shot* | **50%** | **18%** | **31%** | **41%** | **20%** | **33%** |

Exhibit 25 – Semantic similarity by ROUGE (F1 score)

It should be noted that the sentences created by the few-shot learning prompts are very prescriptive and do not deviate at all from the instructions given, and do not deviate much from the ground-truth as highlighted in the semantic similarly by cosine distance. In a practical sense, however, it would be unwise to create an attribution analysis report that contains sentences with the same repetitive structure over and over. One way to overcome this problem is to have a LLM generating variations of the sentences with the correct factor/keyword combination and use these in the prompt as ICL while looping through the sectors. Given that the AI Agent is able to correctly output the sentences as instructed, the end result would be more creative and less dull. Another way is to collect a sizable amount of sentences from actual attribution reports and then fine-tune a model accordingly.

To illustrate, see in Exhibit 26 an extract from one of the outputs generated by the zero-shot prompt, highlighting the diverse ways the LLM explained the effects.

| Allocation | Selection |
| --- | --- |
| The Communication sector was underweighted in the portfolio compared to the benchmark (-0.004), which led to a negative allocation effect (-0.000115). This means that the portfolio's allocation to the Communication sector was less than the benchmark's allocation, which negatively impacted the portfolio's performance. | Despite the underweighting, the Communication sector outperformed the benchmark (0.02), resulting in a positive selection effect (0.001448). This indicates that the specific securities chosen within the Communication sector performed better than those in the benchmark, contributing positively to the portfolio's performance. |
| The Consumer Discret. sector was underweighted in the portfolio compared to the benchmark (-0.0305), which contributed negatively to the portfolio's performance (-0.000459). This suggests that the portfolio could have benefited from a higher allocation to this sector. | The portfolio's return from the Consumer Discret. sector underperformed the benchmark's return (-0.0385), leading to a negative selection effect (-0.004508). This indicates that the specific securities chosen within this sector did not perform as well as those in the benchmark. |
| The Consumer Staples sector was underweighted in the portfolio compared to the benchmark (Portfolio Weight: 0.0158, Benchmark Weight: 0.0526). This underweighting contributed positively to the portfolio's performance (0.004237), as the sector underperformed the overall market. | Despite the underweighting, the stock selection within the Consumer Staples sector underperformed the benchmark (Portfolio Return: -0.0434, Benchmark Return: 0.0169). This resulted in a negative selection effect (-0.000953). |

Exhibit 26 – Zero-shot prompt output

Method 2: Keywords Reasoning Accuracy

Accuracy is measured on two numerical factors (Value) and three keywords (Sector Weight for allocation and Sector Performance for allocation and selection). In general, average accuracy of few-shot learning is in the range of 92-100% and for zero-shot is between 60 and 97%, indicating as expected that few-shot works better than zero-shot.

The LLM is more accurate in describing selection effect than is in describing allocation effect, evidenced by the fact that the "Sector Performance" accuracy is higher on selection than is on allocation. This is not too surprising considering that the reasoning around the allocation effect is more complex than for selection because it requires the right combination of two keywords instead of just one. This is even more pronounced in the zero-shot case, where accuracy is only 60% in the "Sector Performance" keywork for allocation. This result is in accordance with the two semantic similarity scores methods described previously.

Within the few-shot prompts, they all do a similar satisfactory job in 'Value' (on both effects) and 'Sector Weight' (on allocation). On 'Sector Performance' (on allocation), the *few_shot_3* which includes a numerical example demonstrates higher accuracy. Exhibit 27 summarizes these findings.

|  | Allocation Effect | | | Selection Effect | |
| --- | --- | --- | --- | --- | --- |
|  | Value | Sector Weight | Sector Performance | Value | Sector Performance |
| few_1 | 96% | 99% | 92% | 100% | 100% |
| few_2 | 99% | 99% | 94% | 99% | 99% |
| few_3 | 98% | 98% | 97% | 98% | 99% |
| *Few Shot* | *97%* | *99%* | *93%* | *99%* | *100%* |
| zero_2 | 96% | 97% | 60% | 97% | 91% |
| *Zero Shot* | *96%* | *97%* | *60%* | *97%* | *91%* |

Exhibit 27 – Keywords reasoning Accuracy

One feasible way to increase accuracy of the "Sector Performance" would be to create a dynamic feedback loop with the use of the memory construct available within the agent function. The idea is to evaluate the answer and in case the evaluation is negative request the LLM to rerun the prompt query, but without revealing the correct answer. Memory has not been used but it could be helpful in retrieving the previous results and then enriching the subsequent prompt, forcing the agent to restart the process. This dynamic evaluation loop could be a cost-effective way of reducing even further inaccuracies, considering that the process without it has proven to be of high accuracy already. In combining this feedback feature with a more diverse sentence generation function (as discussed previously) an innovative tool could be implemented to help with the written analysis of larger datasets, for example. Exhibit 28 illustrates the process.

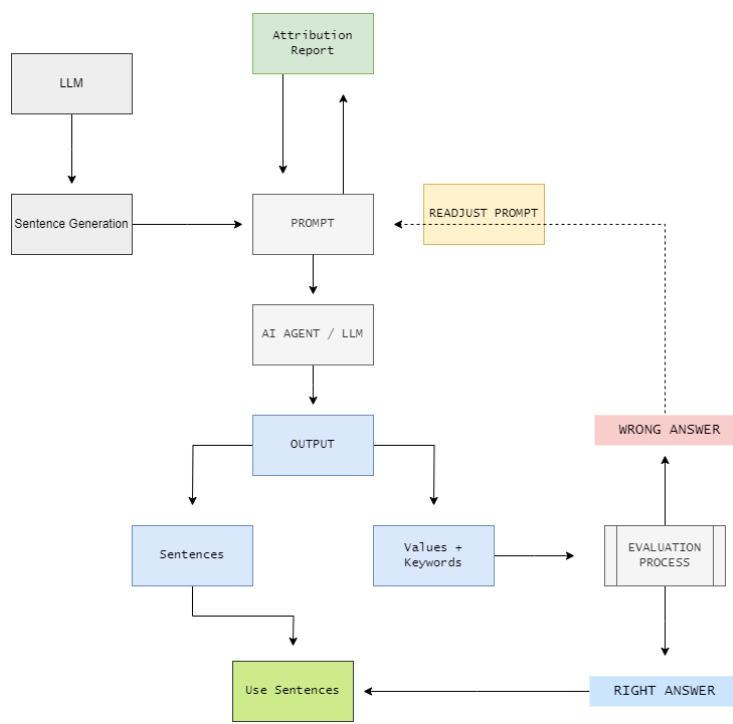

Exhibit 28 – Sentence Generation with Dynamic Evaluation Feedback

Objective #2:

In general, GPT4 did not experienced any issues related to inference time, producing a numerical result, and following output instructions while running the micro level calculations. However, it did experience issues, especially on producing numerical results and following output instructions while running the macro level calculations and some of the question-answers.

Method #1: Calculation of a multi-level attribution report.

In terms of the micro level calculations, GPT4 did not have any issues running the prompt fund by fund, in sequence, in one go. It obtained perfect accuracy in both the sector and parent level calculations – Exhibit 29 summarized the findings. The interesting aspect is that we did not instruct the LLM on how to aggregate the results by the parent type, but it assumed correctly that it was by summation.

|  | Allocation Effect Score | Accuracy | Selection Effect Score | Accuracy |
|---|---|---|---|---|
| **Portfolio Defensive** | **60** | **100%** | **60** | **100%** |
| Sectors | 48 | 100% | 48 | 100% |
| Cyclical | 4 | 100% | 4 | 100% |
| Defensive | 4 | 100% | 4 | 100% |
| Sensitive | 4 | 100% | 4 | 100% |
| **Portfolio Growth** | **60** | **100%** | **60** | **100%** |
| Sectors | 48 | 100% | 48 | 100% |
| Cyclical | 4 | 100% | 4 | 100% |
| Defensive | 4 | 100% | 4 | 100% |
| Sensitive | 4 | 100% | 4 | 100% |
| **Portfolio Value** | **60** | **100%** | **60** | **100%** |
| Sectors | 48 | 100% | 48 | 100% |
| Cyclical | 4 | 100% | 4 | 100% |
| Defensive | 4 | 100% | 4 | 100% |
| Sensitive | 4 | 100% | 4 | 100% |
| **Total Sectors** | **144** | **100%** | **144** | **100%** |
| **Total Types** | **36** | **100%** | **36** | **100%** |

Exhibit 29 – Micro level calculations accuracy

In terms of the macro level calculations, GPT4 did have several issues while running the prompt fund by fund, in sequence. It eventually was able to obtain perfect accuracy, see results in Exhibit 30, but not before several attempts, especially on Portfolio Growth and Portfolio Value. As for the Portfolio Defensive, it was able to run in sequence with perfect accuracy from the first try. We are not sure of the reasons for this performance discrepancy. It could be related to the fact that the prompt is complicated with several formulas and instructions, and it may need numerical examples in the prompt. Exhibit 31 illustrates how accuracy is impacted when the LLM was not able to correctly follow the instructions.

|  | Allocation Effect Score | Accuracy | Selection Effect Score | Accuracy |
|---|---|---|---|---|
| **Portfolio Defensive** | **12** | **100%** | **12** | **100%** |
| Cyclical | 4 | 100% | 4 | 100% |
| Defensive | 4 | 100% | 4 | 100% |
| Sensitive | 4 | 100% | 4 | 100% |
| **Portfolio Growth** | **12** | **100%** | **12** | **100%** |
| Cyclical | 4 | 100% | 4 | 100% |
| Defensive | 4 | 100% | 4 | 100% |
| Sensitive | 4 | 100% | 4 | 100% |
| **Portfolio Value** | **12** | **100%** | **12** | **100%** |
| Cyclical | 4 | 100% | 4 | 100% |
| Defensive | 4 | 100% | 4 | 100% |
| Sensitive | 4 | 100% | 4 | 100% |
| **Grand Total** | **36** | **100%** | **36** | **100%** |

Exhibit 30 – Macro level calculations with perfect accuracy

|  | Allocation Effect Score | Accuracy | Selection Effect Score | Accuracy |
|---|---|---|---|---|
| **Portfolio Defensive** | **12** | **100%** | **12** | **100%** |
| Cyclical | 4 | 100% | 4 | 100% |
| Defensive | 4 | 100% | 4 | 100% |
| Sensitive | 4 | 100% | 4 | 100% |
| **Portfolio Growth** | **9** | **75%** | **9** | **75%** |
| Cyclical | 3 | 75% | 3 | 75% |
| Defensive | 3 | 75% | 3 | 75% |
| Sensitive | 3 | 75% | 3 | 75% |
| **Portfolio Value** | **7** | **58%** | **6** | **50%** |
| Cyclical | 2 | 50% | 2 | 50% |
| Defensive | 3 | 75% | 2 | 50% |
| Sensitive | 2 | 50% | 2 | 50% |
| **Grand Total** | **28** | **78%** | **27** | **75%** |

Exhibit 31 – Macro level calculations with lower accuracy

As described in Exhibit 20, the prompts used for the micro and macro attribution calculations contained several mathematical formulas and step by step calculation instructions. Despite some difficulties especially on the macro attribution calculations, it is a remarkable feature of GPT-4 to be able to correctly follow instructions in sequence and to translate the formulae into pandas functions. GPT-3.5 was unable to perform any of these calculations, for instance. Taking this into consideration, it may not be too difficult to extent the agent capabilities to more complex calculations such as the use of geometric returns, multi-period, and multi-currency attribution analysis, and use it in more complex portfolios like fixed income, funds with derivatives and short positions.

Method #2: Question Answer (QA)

Two types of QA was run for a total of 140 questions: 70 QA with multiple choice (QAMC) and 70 QA with a direct command to calculate a specific effect of a specific sector (QACalc). Accuracy obtained was greater than 84%, which is a robust performance and could be a potential passing score in an examination. Exhibit 32 summarizes the findings.

However, this is also an indication that the agent is better in calculating results for the overall portfolio at once as shown in Exhibit 29 in the aforementioned method #1 at the micro level. One way to take advantage of this feature and make the agent be more accurate in a QA set-up would be to firstly instruct it to calculate a complete attribution report and then secondly request it to answer the question.

| Question Type | Correct | Total Questions | Accuracy |
|---|---|---|---|
| QAMC | 62 | 70 | 89% |
| QACalc | 59 | 70 | 84% |

Exhibit 32 – QA accuracy

Out of the 140 answers, GPT-4 did correctly follow the output instructions 139 times, failing just once when it answered the question repeating the question with all the multiple choice and concluded: "Given the calculated allocation effect, the closest value is D: -0.00046%."

6.   Conclusion

The main hypothesis underlying this research was to investigate whether an AI agent running in a suitable foundation model with proper prompt engineering could with minimum error accomplish two objectives: i) produce a performance attribution analysis and ii) calculate its main components following specific formulae and step by step instructions. In light of the encouraging results obtained on both of these objectives, we conclude that the research is successful, and contributes to the field of prompt engineering, AI Agents, and evaluation.

On objective #1, from a complete portfolio attribution report, the agent was able to correctly reason on the main drivers of allocation and selection effect while following exactly the output instructions. It is clear that it struggled more on the allocation effect reasoning than on selection but achieving 93% keyword average reasoning accuracy is still an incredible feature, thus opening the process for further improvement, especially around the use of memory to implement an innovative dynamic evaluation feedback loop process, and fine-tuning.

On objective #2, from basic portfolio information the agent was able to correctly perform macro and micro attribution analysis. Moreover, given the complex nature of some of the output instructions and formulae provided, this research highlights the remarkable feature of using state-of-the-art models like GPT-4 as the main orchestrator of an AI agent. This feature was clearly evidenced in the question-answer portion where the agent scored correctly more than 84% of the time in what could be a potential passing score in an official examination. In addition, the results suggest that more complex analyses could be performed such as multi-temporal, multi-currency, and the analysis of asset classes that feature more attribution effects like for fixed income portfolios.

In conclusion, we believe that the reassuring results presented in this research are a clear signal that the fields of performance attribution analysis and investment performance appraisal are ripe for disruption. We have only touched the most basic of attribution models and given the continuous innovation in the field of AI agents, there is still much more to be accomplished.

7.     Future Work

We highlight below some of the areas for future work related to the use of AI agents in the field of performance attribution analysis and performance appraisal:

- Test other LLMs, although GPT-3.5 was unable to perform some of the more complex instructions
- Test different types of prompts, with the use of external tools and functions calling features
- Adjust the agent python class to include an intermediate step that verifies the results and re-runs the agent - dynamic evaluation feedback loop process.
- Ground the attribution analysis using knowledge graph
- Calculate multi-period, multi-currency, geometric return, fixed income attribution analysis, use of shorts and derivatives.

## 8. Reference

Allen, Gregory C. 1991. "Performance Attribution for Global Equity Portfolios." Journal of Portfolio Management 18 (1): 59–65.

Bacon, Carl R. 2002. "Excess Returns—Arithmetic or Geometric?" Journal of Performance Measurement 6 (3): 23–31.

Bacon, Carl R. 2008. "Practical Portfolio Performance Measurement and Attribution". John Wiley & Sons.

Bacon, Carl. 2019. "Performance Attribution History and Progress". CFA Institute Research Foundation.

Bacon, Carl R., CIPM, and Marc A. Wright, CFA. 2019. "CIPM Reading 5 Return Attribution". CFA Institute.

Brinson, Gary P., and Nimrod Fachler. 1985. "Measuring Non-US Equity Portfolio Performance." Journal of Portfolio Management 11 (3): 73–76.

Brinson, Gary P., L. Randolph Hood, Jr., and Gilbert L. Beebower. 1986. "Determinants of Portfolio Performance." Financial Analysts Journal 42 (4): 39–44.

Carino, David R. 1999. "Combining Attribution Effects over Time." Journal of Performance Measurement 3 (4): 5–14.

Chase, H. 2022. https://python.langchain.com/docs/modules/agents/

Dai, Tianci, and Mark Elliot. 2014/2015. "Fixed Income Attribution with Carry Effect." Journal of Performance Measurement 19 (2): 7–18.

Dietz, Peter O. 1966. "Pension Funds: Measuring Investment Performance". Graduate School of Business of Columbia University and Free Press.

Dong, Q., L. Li, D. Dai, et al. 2023. "A survey for in-context learning". CoRR, abs/2301.00234.

Lin, C-Y. 2004. "ROUGE: A Package for Automatic Evaluation of Summaries". Text Summarization Branches Out, Proceedings of the ACL-04 Workshop.

McLaren, Andrew. 2002. "A Framework for Multiple Currency Fixed Income Attribution." Journal of Performance Measurement 6 (4): 59–79.

Milvus. (2023). https://zilliz.com/what-is-milvus.

Patil, Shishir et al. 2023. "Gorilla: Large Language Model Connected with Massive APIs" arXiv preprint arXiv:2305.15334.

Shen et al. 2023. "HuggingGPT: Solving AI Tasks with ChatGPT and its Friends in HuggingFace" arXiv preprint arXiv:2303.17580.

Wang, X., J. Wei, D. Schuurmans, et al. 2023. "Self-consistency improves chain of thought reasoning in language models." In The Eleventh International Conference on Learning Representations, ICLR 2023, Kigali, Rwanda, May 1-5.

Weng, Lilian. (Jun 2023). "LLM-powered Autonomous Agents". Lil'Log. https://lilianweng.github.io/posts/2023-06-23-agent/.

Wei et al. "Chain of thought prompting elicits reasoning in large language models." NeurIPS 2022.

Yao et al. "ReAct: Synergizing reasoning and acting in language models." ICLR 2023.

Yao et al. "Tree of Thoughts: Deliberate Problem Solving with Large Language Models." arXiv preprint arXiv:2305.10601.

Zhiheng Xi et al. 2023. "The Rise and Potential of Large Language Model Based Agents: A Survey". arXiv preprint arXiv:2309.07864.

# 9. Appendix

## Objective #1: Sample Attribution Report

| GICS Sector | Portfolio Weight | Benchmark Weight | Portfolio Return | Benchmark Return | Variation in Weight | Variation in Return | Allocation Effect | Selection Effect | Total Contribution | Period | Fund | Benchmark |
|---|---|---|---|---|---|---|---|---|---|---|---|---|
| Energy | 0.0370 | 0.0750 | 0.0320 | 0.0430 | -0.0380 | -0.0110 | 0.0011 | -0.0004 | 0.0007 | 1/31/2022 to 3/31/2022 | Portfolio Defensive | Benchmark Defensive |
| Materials | 0.0530 | 0.0890 | 0.0470 | 0.0520 | -0.0360 | -0.0050 | 0.0007 | -0.0003 | 0.0004 | 1/31/2022 to 3/31/2022 | Portfolio Defensive | Benchmark Defensive |
| Industrials | 0.0680 | 0.0930 | 0.0570 | 0.0610 | -0.0250 | -0.0040 | 0.0003 | -0.0003 | 0.0000 | 1/31/2022 to 3/31/2022 | Portfolio Defensive | Benchmark Defensive |
| Consumer Discret. | 0.0890 | 0.0930 | 0.0940 | 0.0730 | -0.0040 | 0.0210 | 0.0000 | 0.0019 | 0.0019 | 1/31/2022 to 3/31/2022 | Portfolio Defensive | Benchmark Defensive |
| Consumer Staples | 0.1450 | 0.0880 | 0.1130 | 0.0970 | 0.0570 | 0.0160 | 0.0015 | 0.0023 | 0.0038 | 1/31/2022 to 3/31/2022 | Portfolio Defensive | Benchmark Defensive |
| Health Care | 0.1160 | 0.0940 | 0.0710 | 0.0770 | 0.0220 | -0.0060 | 0.0001 | -0.0007 | -0.0006 | 1/31/2022 to 3/31/2022 | Portfolio Defensive | Benchmark Defensive |
| Financials | 0.0770 | 0.1360 | 0.0430 | 0.0570 | -0.0590 | -0.0140 | 0.0009 | -0.0011 | -0.0002 | 1/31/2022 to 3/31/2022 | Portfolio Defensive | Benchmark Defensive |
| IT | 0.0860 | 0.1120 | 0.1060 | 0.0820 | -0.0260 | 0.0240 | -0.0003 | 0.0021 | 0.0018 | 1/31/2022 to 3/31/2022 | Portfolio Defensive | Benchmark Defensive |
| Communication | 0.0680 | 0.0810 | 0.0930 | 0.1030 | -0.0130 | -0.0100 | -0.0004 | -0.0007 | -0.0011 | 1/31/2022 to 3/31/2022 | Portfolio Defensive | Benchmark Defensive |
| Utilities | 0.1380 | 0.0720 | 0.0890 | 0.0670 | 0.0660 | 0.0220 | -0.0003 | 0.0030 | 0.0027 | 1/31/2022 to 3/31/2022 | Portfolio Defensive | Benchmark Defensive |
| Real Estate | 0.1030 | 0.0670 | 0.0940 | 0.0790 | 0.0360 | 0.0150 | 0.0003 | 0.0015 | 0.0018 | 1/31/2022 to 3/31/2022 | Portfolio Defensive | Benchmark Defensive |
| Cash | 0.0200 | 0.0000 | 0.0038 | 0.0000 | 0.0200 | 0.0038 | -0.0014 | 0.0001 | -0.0014 | 1/31/2022 to 3/31/2022 | Portfolio Defensive | Benchmark Defensive |
| Energy | 0.0450 | 0.0770 | 0.0100 | 0.0150 | -0.0320 | -0.0050 | -0.0006 | -0.0002 | -0.0009 | 4/1/2022 to 6/30/2022 | Portfolio Defensive | Benchmark Defensive |
| Materials | 0.0480 | 0.0920 | -0.0200 | -0.0300 | -0.0440 | 0.0100 | 0.0011 | 0.0005 | 0.0016 | 4/1/2022 to 6/30/2022 | Portfolio Defensive | Benchmark Defensive |
| Industrials | 0.0600 | 0.0890 | -0.0050 | 0.0150 | -0.0290 | -0.0200 | -0.0006 | -0.0012 | -0.0018 | 4/1/2022 to 6/30/2022 | Portfolio Defensive | Benchmark Defensive |
| Consumer Discret. | 0.0920 | 0.0910 | -0.0300 | -0.0190 | 0.0010 | -0.0110 | 0.0000 | -0.0010 | -0.0010 | 4/1/2022 to 6/30/2022 | Portfolio Defensive | Benchmark Defensive |
| Consumer Staples | 0.1200 | 0.0910 | 0.0250 | 0.0200 | 0.0290 | 0.0050 | 0.0007 | 0.0006 | 0.0013 | 4/1/2022 to 6/30/2022 | Portfolio Defensive | Benchmark Defensive |
| Health Care | 0.1400 | 0.0920 | 0.0110 | 0.0150 | 0.0480 | -0.0040 | 0.0010 | -0.0006 | 0.0004 | 4/1/2022 to 6/30/2022 | Portfolio Defensive | Benchmark Defensive |
| Financials | 0.0700 | 0.1340 | -0.0300 | -0.0400 | -0.0640 | 0.0100 | 0.0022 | 0.0007 | 0.0029 | 4/1/2022 to 6/30/2022 | Portfolio Defensive | Benchmark Defensive |
| IT | 0.0900 | 0.1160 | -0.0500 | -0.0150 | -0.0260 | -0.0350 | 0.0003 | -0.0032 | -0.0029 | 4/1/2022 to 6/30/2022 | Portfolio Defensive | Benchmark Defensive |
| Communication | 0.0720 | 0.0740 | -0.0350 | -0.0250 | -0.0020 | -0.0100 | 0.0000 | -0.0007 | -0.0007 | 4/1/2022 to 6/30/2022 | Portfolio Defensive | Benchmark Defensive |
| Utilities | 0.1550 | 0.0790 | 0.0110 | 0.0200 | 0.0760 | -0.0090 | 0.0019 | -0.0014 | 0.0005 | 4/1/2022 to 6/30/2022 | Portfolio Defensive | Benchmark Defensive |
| Real Estate | 0.0980 | 0.0650 | 0.0030 | 0.0150 | 0.0330 | -0.0120 | 0.0007 | -0.0012 | -0.0005 | 4/1/2022 to 6/30/2022 | Portfolio Defensive | Benchmark Defensive |
| Cash | 0.0100 | 0.0000 | 0.0038 | 0.0000 | 0.0100 | 0.0038 | 0.0001 | 0.0000 | 0.0001 | 4/1/2022 to 6/30/2022 | Portfolio Defensive | Benchmark Defensive |

## Objective #2: Sample Attribution Report

| GICS Type | GICS Sector | Portfolio Weight | Benchmark Weight | Portfolio Return | Benchmark Return | Period | Fund | Benchmark |
|---|---|---|---|---|---|---|---|---|
| Sensitive | Energy | 0.0370 | 0.0750 | 0.0320 | 0.0430 | 1/31/2022 to 3/31/2022 | Portfolio Defensive | Benchmark Defensive |
| Cyclical | Materials | 0.0530 | 0.0890 | 0.0470 | 0.0520 | 1/31/2022 to 3/31/2022 | Portfolio Defensive | Benchmark Defensive |
| Sensitive | Industrials | 0.0680 | 0.0930 | 0.0570 | 0.0610 | 1/31/2022 to 3/31/2022 | Portfolio Defensive | Benchmark Defensive |
| Cyclical | Consumer Discret. | 0.0890 | 0.0930 | 0.0940 | 0.0730 | 1/31/2022 to 3/31/2022 | Portfolio Defensive | Benchmark Defensive |
| Defensive | Consumer Staples | 0.1450 | 0.0880 | 0.1130 | 0.0970 | 1/31/2022 to 3/31/2022 | Portfolio Defensive | Benchmark Defensive |
| Defensive | Health Care | 0.1160 | 0.0940 | 0.0710 | 0.0770 | 1/31/2022 to 3/31/2022 | Portfolio Defensive | Benchmark Defensive |
| Cyclical | Financials | 0.0770 | 0.1360 | 0.0430 | 0.0570 | 1/31/2022 to 3/31/2022 | Portfolio Defensive | Benchmark Defensive |
| Sensitive | IT | 0.0860 | 0.1120 | 0.1060 | 0.0820 | 1/31/2022 to 3/31/2022 | Portfolio Defensive | Benchmark Defensive |
| Sensitive | Communication | 0.0680 | 0.0810 | 0.0930 | 0.1030 | 1/31/2022 to 3/31/2022 | Portfolio Defensive | Benchmark Defensive |
| Defensive | Utilities | 0.1380 | 0.0720 | 0.0890 | 0.0670 | 1/31/2022 to 3/31/2022 | Portfolio Defensive | Benchmark Defensive |
| Cyclical | Real Estate | 0.1030 | 0.0670 | 0.0940 | 0.0790 | 1/31/2022 to 3/31/2022 | Portfolio Defensive | Benchmark Defensive |
| Defensive | Cash | 0.0200 | 0.0000 | 0.0038 | 0.0000 | 1/31/2022 to 3/31/2022 | Portfolio Defensive | Benchmark Defensive |
| Sensitive | Energy | 0.0450 | 0.0770 | 0.0100 | 0.0150 | 4/1/2022 to 6/30/2022 | Portfolio Defensive | Benchmark Defensive |
| Cyclical | Materials | 0.0480 | 0.0920 | -0.0200 | -0.0300 | 4/1/2022 to 6/30/2022 | Portfolio Defensive | Benchmark Defensive |
| Sensitive | Industrials | 0.0600 | 0.0890 | -0.0050 | 0.0150 | 4/1/2022 to 6/30/2022 | Portfolio Defensive | Benchmark Defensive |
| Cyclical | Consumer Discret. | 0.0920 | 0.0910 | -0.0300 | -0.0190 | 4/1/2022 to 6/30/2022 | Portfolio Defensive | Benchmark Defensive |
| Defensive | Consumer Staples | 0.1200 | 0.0910 | 0.0250 | 0.0200 | 4/1/2022 to 6/30/2022 | Portfolio Defensive | Benchmark Defensive |
| Defensive | Health Care | 0.1400 | 0.0920 | 0.0110 | 0.0150 | 4/1/2022 to 6/30/2022 | Portfolio Defensive | Benchmark Defensive |
| Cyclical | Financials | 0.0700 | 0.1340 | -0.0300 | -0.0400 | 4/1/2022 to 6/30/2022 | Portfolio Defensive | Benchmark Defensive |
| Sensitive | IT | 0.0900 | 0.1160 | -0.0500 | -0.0150 | 4/1/2022 to 6/30/2022 | Portfolio Defensive | Benchmark Defensive |
| Sensitive | Communication | 0.0720 | 0.0740 | -0.0350 | -0.0250 | 4/1/2022 to 6/30/2022 | Portfolio Defensive | Benchmark Defensive |
| Defensive | Utilities | 0.1550 | 0.0790 | 0.0110 | 0.0200 | 4/1/2022 to 6/30/2022 | Portfolio Defensive | Benchmark Defensive |
| Cyclical | Real Estate | 0.0980 | 0.0650 | 0.0030 | 0.0150 | 4/1/2022 to 6/30/2022 | Portfolio Defensive | Benchmark Defensive |
| Defensive | Cash | 0.0100 | 0.0000 | 0.0038 | 0.0000 | 4/1/2022 to 6/30/2022 | Portfolio Defensive | Benchmark Defensive |